# Ferromagnetic bubble clusters in $Y_{0.67}Ca_{0.33}MnO_3$ thin films

Jeehoon Kim,[1] N. Haberkorn,[2] L. Civale,[1] P. C. Dowden[1] and R. Movshovich.[1]

[1] Los Alamos National Laboratory, Los Alamos, NM 87545.

[2] Centro Atómico Bariloche, 8400 Bariloche, Argentina.

(Dated: 3 January 2013)

We studied the ferromagnetic topology in a $Y_{0.67}Ca_{0.33}MnO_3$ thin film with a combination of magnetic force microscopy and magnetization measurements. Our results show that the spin-glass like behavior, reported previously for this system, could be attributed to frustrated interfaces of the ferromagnetic clusters embedded in a non-ferromagnetic matrix. We found temperature dependent changes of the magnetic topology at low temperatures, which suggests a non-static $Mn^{3+}/Mn^{4+}$ ratio.

The coexistence of distinct magnetic phases within a single sample of a perovskite manganite compound ($A_{1-x}B_xMnO_3$: A and B represent rare-earth and alkaline-earth elements) has been intensely studied for decades because of both technological applications and fascinating physics.[1,2,3,4] The electronic and magnetic properties of manganites can be tuned by substitution of cations and/or by the modification of the oxygen content. A certain range of doping levels results in a drastic change of resistance, i.e., a colossal magnetoresistance (CMR) effect.[5] These properties are strongly related to a structural distortion, which can be analyzed by the tolerance factor $t$, defined as $t = (<r_{A/B} + r_O>)/(<r_{Mn} + r_O>)$, where $r_{A/B}$, $r_O$, and $r_{Mn}$ are the radii of the rare-earth/alkaline-earth, the oxide, and the manganese, respectively.[6] The structure of perovskite materials is, in general, stable in the range of $0.8 < t < 1$. The perovskite structure tends to distort when $t$ deviates from 1 and, in particular, manganites do not undergo a metal-insulator transition (MIT) when $t < 0.91$. In this extreme regime $Gd_{2/3}Ca_{1/3}MnO_3$ (GCMO, $t \approx 0.89$) and $Y_{2/3}Ca_{1/3}MnO_3$ (YCMO, $t \approx 0.88$) appear. Both systems display a ferromagnetic (FM) ordering at around 80 K, associated with the Mn ions, and none of them exhibit a MIT.[7,8] Recently we reported magnetic phase coexistence and magnetization reversal in ferrimagnetic GCMO thin films,[9] which is consistent with those previously reported in single crystals.[10] The magnetic response of YCMO is governed by short-range FM correlations below the Curie temperature ($T_C$), and its magnetization shows magnetic history dependence.[11] The YCMO system shows a spin glass-like behavior with a freezing temperature ($T_f$) of about 30 K. The dynamics above $T_f$ is attributed to a thermally activated redistribution of FM-ordered clusters and a random dipolar interaction of their magnetic moments.[8]

Although magnetic properties might be expected to be similar for GCMO and YCMO owing to similar structural distortions and tolerance factor, the resulting magnetic properties are in fact different. In GCMO the magnetic coupling via a $d$-$f$ exchange interaction between Gd and Mn plays an important role and leads to the presence of a compensation temperature ($T_{comp}$) as a result of a competing ferrimagnetic order and a giant magnetostriction.[12] YCMO, on the other hand, displays a spin-glass behavior, which can be interpreted in terms of FM clusters with an associated lattice distortion and magnetic inhomogeneity of the system.[13,14] In this Letter we report the FM phase topology of a YCMO thin film studied by magnetic force microscopy (MFM). We observe FM nanoclusters embedded in a non-FM matrix, in support of the previously reported data and interpretations.[8,11,13] Images of the FM nanoclusters as a function of temperature demonstrate that the nanoclusters exhibit fluctuations under specific conditions.

The $Y_{0.67}Ca_{0.33}MnO_3$ (YCMO) thin film was grown by pulsed-laser deposition (PLD) on a $SrTiO_3$ (100) substrate using a commercial target with the same chemical composition. The substrate temperature was kept at 790 °C in an oxygen atmosphere at a pressure of 200 mTorr. After deposition, the $O_2$ pressure was increased up to 200 Torr, and the temperature was decreased down to room temperature at a rate of 30 °C/min. Bulk YCMO is an orthorhombic perovskite with lattice parameters of $a/\sqrt{2} = 0.392$ nm, $b/2 = 0.375$ nm, $c/\sqrt{2} = 0.372$ nm.[8] The YCMO film was examined by x-ray diffractometry, and was found to be single phase with a (0$l$0) orientation. The lattice parameters of the film [$a/\sqrt{2} = 0.392$ (1), $b/2 = 0.378$ (1), $c/\sqrt{2} = 0.374(1)$] were determined using (0$l$0), (200), and (002) reflections from a four-circle diffractometer/goniometer. No additional peaks due to secondary phases or different crystalline orientations were observed (see figure 1). The rocking curve FWHM of the (040) peak of the film was 0.24°. Furthermore, the four peaks at 90° intervals in the $\phi$ scan make evident the existence of in-plane order of the film. The film thickness of 33(2) nm was determined by a low-angle x-ray reflectivity measurement with an angular resolution of 0.005°.

A Quantum Design MPMS superconducting quantum interference device (SQUID) magnetometer was used for measurements of the global magnetization with the magnetic field perpendicular to the film surface. All MFM measurements described in this paper were carried out in a home-built low-temperature MFM apparatus.[15] MFM images were obtained in high vacuum of $1\times10^{-6}$ Torr in a frequency-modulated mode. Commercially available cantilevers with a Co/Cr coating layer[16] were used for MFM measurements. The MFM tip was magnetized along the tip axis in a field of 3 T prior to MFM measurements. The external magnetic field ($H$) was always applied perpendicular to the film surface. The negative frequency shift of the tip results from the attractive interaction between the tip and the sample magnetization. Therefore, the dark features in the MFM image, displaying a negative frequency shift of the tip, indicate that the sample magnetization is parallel to the tip magnetization.

In figure 2(a) we present magnetization ($M$) vs Temperature ($T$) at $\mu_0H = 0.1$ T for **H** perpendicular to the surface. The global magnetic measurements were performed in the same configuration as the local measurements in MFM. The $M$-$T$ curve shows an inflection at approximately 75 K, which corresponds to the FM order reported previously for a bulk sample.[8] Figure 2(b) shows the coercive field ($H_c$) vs $T$ obtained from magnetic hysteresis loops at each temperature (see inset). The data show an increase of $H_c$ below 30 K, which corresponds to the freezing temperature, signaling that the system possibly undergoes a spin-glass transition.[8] Additionally, the saturation magnetization ($M_s$), obtained from the subtraction of the paramagnetic background, is always smaller than the theoretical value of $M_s \approx 560$ emu/cm$^3$. The $M_s$ value at 5 K is $170 \pm 30$ emu/cm$^3$, indicating the presence of non-FM regions or frustrated magnetism.

Figures 3(a)-3(e) display MFM images obtained sequentially at 4 K along an upper branch of the magnetic hysteresis loop after saturation at $\mu_0 H = 1$ T. The MFM image obtained at $\mu_0 H = 0.5$ T [see Fig. 3(a)] shows coexistence of isolated round and elongated domains. The bright domains are antiparallel to the tip magnetization, and their size is around 200 nm. The remanent state ($H=0$), shown in Fig. 3(b), is characterized by dark spots (bubble domains parallel to the tip field), appearing in the matrix of a homogeneous magnetization. The size of the ferromagnetic bubble domains is around 100 nm, smaller than that for $\mu_0 H = 0.5$ T. The shape of the bubbles persists up to $\mu_0 H = -0.1$ T, and changes back to large domains at $\mu_0 H = -0.5$ T, showing similar shapes to those in Fig. 3(a): the cross correlation map, shown in Fig. 3(f), shows strong positive correlations. This indicates that magnetization reversal takes place via rotation of the magnetic domains instead of the nucleation of the reversed domains that expand with increasing $H$. This type of magnetization reversal via domain rotation is a typical signature of phase separated magnetic materials. Data taken at $\mu_0 H = -3$ T (not shown) are similar to those at $\mu_0 H = -1$ T, indicating the saturation of the sample and the presence of non-FM regions in the film. The rapid change of the domain features at low field is related to the stiffness of the magnetic domains due to the dominant shape anisotropy. The out-of plane magnetic saturation field ($H_s$) can be estimated by considering the theoretical expression for an isolated bubble domain, assuming a disk with a diameter of 200 nm and a thickness of 33 nm, $4\pi(1-D)M_s \approx 2000$ Oe, where $D$ is the demagnetization factor[17] and $M_s$ is the saturation magnetization. The experimental values of $H_s$ [see the inset in Fig 2(b)] are around 3600 Oe, larger than the theoretical value of 2000 Oe, indicating that the system presents large domains produced by interconnected bubbles with common boundaries, which is in good agreement with the experimental data [see Fig. 3(e)]. The magnetic topology of YCMO shows isolated bubble domains, different from the topology of the GCMO film,[9] which showed non-symmetric and larger magnetic regions. The presence of the unconnected bubble domains at low field in YCMO can be understood by the inhomogeneous $Mn^{3+}/Mn^{4+}$ ratio as a mechanism for stress relaxation.[18] Figure 4 shows MFM images, obtained sequentially from the same place at 4 K and 10 K, respectively. Thermal drift was negligible from 4 K to 15 K due to the rigid design of the microscope.[9, 15] The images were obtained at $\mu_0 H = -0.1$ T after the sample was saturated at $\mu_0 H = 1$ T, as in Fig. 3(c). The features between 4 K and 10 K have no spatial correlation,[10] indicating that the bubble domains change drastically with the temperature under these conditions.

Having both the MFM images in Figs. 3 and 4 and those discussed in Ref. [8] in GCMO films allows us to discuss similarities and differences between GCMO and YCMO. GCMO thin films exhibit phase coexistence between ferrimagnetic domains and non-ferrimagnetic regions[9] and show larger domains than do YCMO films. The main contrast between the two films arises from the *Gd-Mn* interaction in GCMO, which modifies the magnetism and results in the $T_{comp}$, where the magnetizations from Gd and Mn sublattices are antiparallel and equal to each other. The large changes and the complex behavior of the magnetism due to the *Gd-Mn* antiferromagnetic coupling make the analysis of the evolution of the magnetic domains difficult.[12] There are several possible mechanisms of the magnetic interaction within an assembly of magnetic particles.[19] In general, the dipole-dipole interaction between particles is of primary importance for such systems. A direct exchange interaction via the surface of the bubble domains should be taken into account as well when the clusters are in close contact with each other. Another possible explanation is the presence of frustrated interfaces between FM and non-FM regions, which is also consistent with the spin glass-like behavior reported in the bulk samples.[8] Unconventional glass-like behaviors appear either in bulk manganites with phase separation or in films and multilayers with strained interfaces.[20,21,22,23] No correlation was detected between the FM domains at 4 K and 10 K in this study, which is consistent with random nucleation, and suggests that the intrinsic distortion, due to the low tolerance factor, plays a salient role in the magnetic topology. Although magnetic properties in thin films could be strongly affected by stress and strain,[24] the drastic change of the $H_c(T)$ and the magnetic topology of isolated bubbles at low temperatures (see Fig. 4) suggest a non-static $Mn^{3+}/Mn^{4+}$ ratio as a mechanism for strain relaxation. We believe that the freezing temperature could be associated with changes of the non-FM matrix, which prevent mobility of the FM cluster and produce changes of the dynamics of the material.[8] This hypothesis is in agreement with the fact that $H_s$ does not change significantly between 5 and 30 K (not shown), which indicates no significant change of the demagnetization factor due to coupling between the bubbles.

In conclusion, we studied topology of the magnetic domains in a high quality epitaxial YCMO thin film. Our results show the phase coexistence between FM and non-FM domains and a spin-glass behavior below $T_C$, which is supported by a strong suppression of the saturation magnetization. We found that at low magnetic field the unusually small size of the isolated bubbles. Our observations are consistent with the fluctuation of the $Mn^{3+}/Mn^{4+}$ ratio as the mechanism of strain relaxation in YCMO films. Our temperature-dependent studies show a direct evidence of the spin glass-like behavior and magnetism reported previously in bulk samples. The smaller size of the round shape domains in YCMO, compared to those in GCMO, suggests a potential application for a magnetic memory device, and magnetic template of magnetic pinning centers in superconductors.

We thank J. O. Willis for providing useful comments. This work was supported by the US Department of Energy, Office of Basic Energy Sciences, Division of Materials Sciences and Engineering and the Los Alamos National Laboratory's Laboratory Directed Research and Development Program, Project No. 20130285ER. N.H. is a member of CONICET (Argentina).

FIG. 1. X-ray diffractogram (logarithmic intensity scale) of the YCMO film at room temperature. The inset shows the rocking curve for the (040) reflection.

FIG. 2. (Color online) (a) Magnetization vs temperature at $\mu_0 H =$ 0.1 T. (b) Coercive field vs temperature obtained from magnetic hysteresis loops. Inset: typical hysteresis loop at 15 K. All the measurements were performed with H $\perp$ to the surface.

FIG. 3. (Color online) (a)-(e) MFM images in YCMO taken sequentially in different fields at 4 K. (f) Cross correlation images between (a) and (d). The bright spot marked by the white arrow shows a strong positive correlation and indicates a similar domain structure between (a) and (d). The position of the spot in the correlation map is off-centered, indicating a small field drift is present. The tip lift height was 100 nm from the surface.

FIG. 4. (Color online) MFM images obtained at different temperatures. (a) The image was obtained in $\mu_0 H =$ -0.1 T after the sample was saturated in the field of 1 T. (b) The MFM image taken and after warming the sample of (a) in the same field of -0.1 T. The tip lift height was 100 nm above the sample surface. No spatial correlation was resolved between 4 K and 10 K.

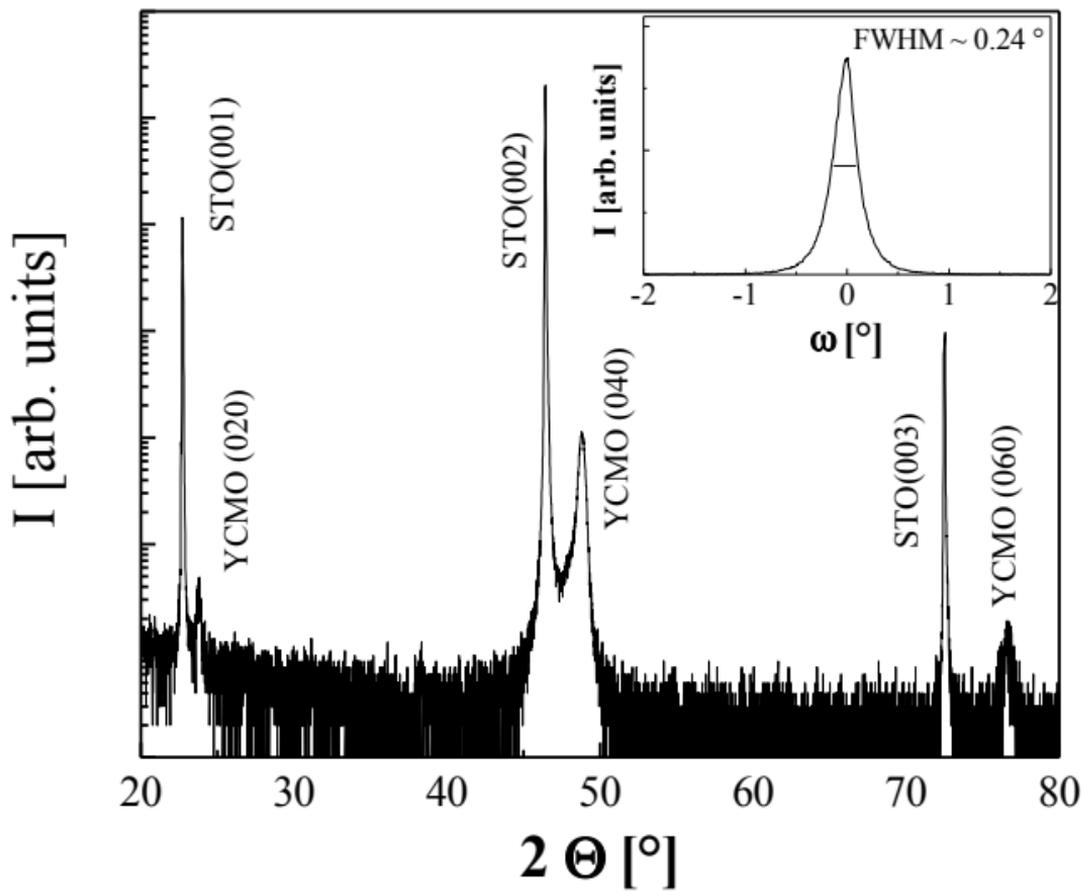

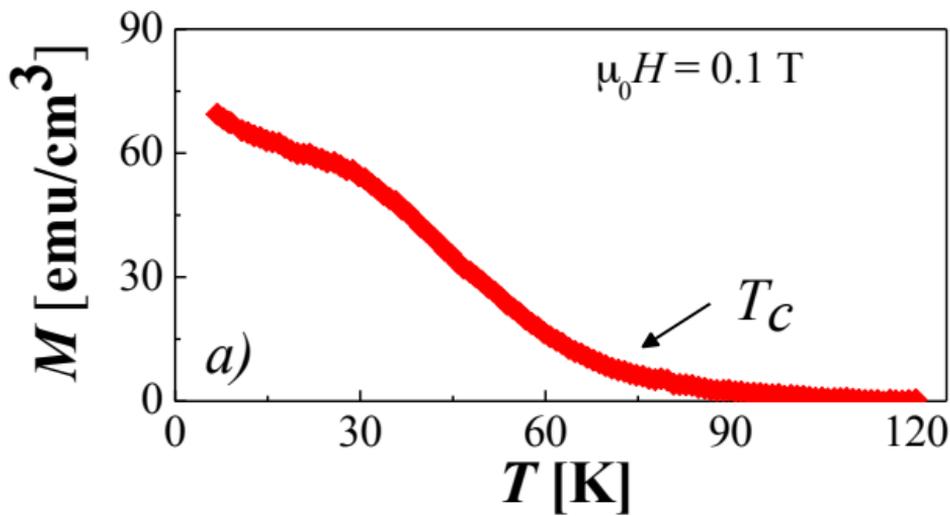
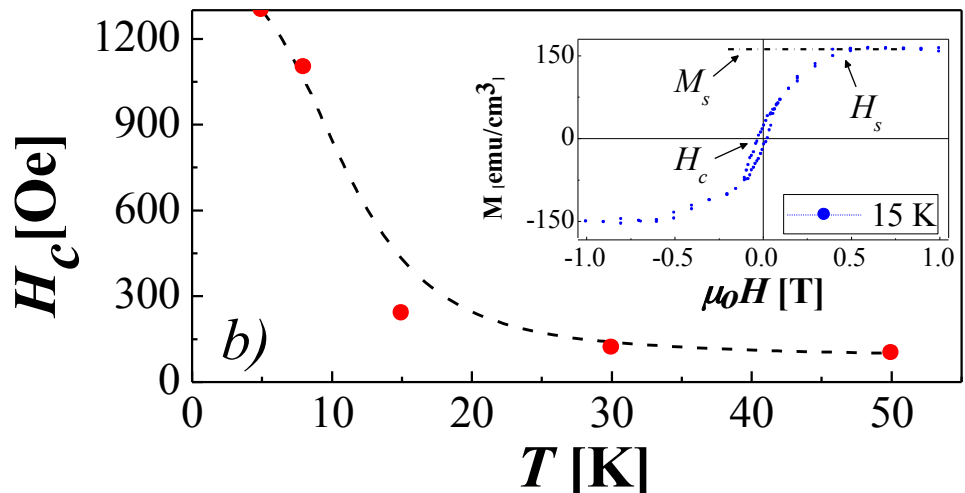

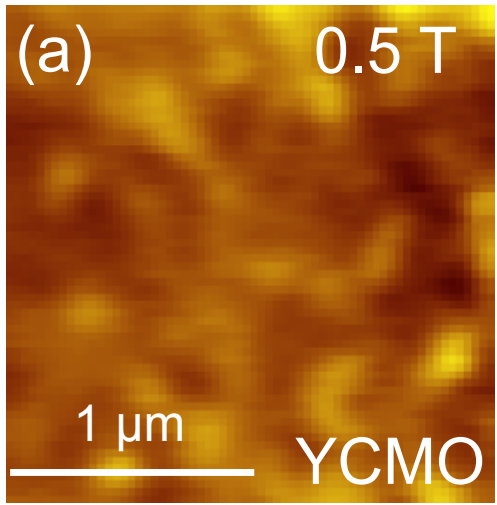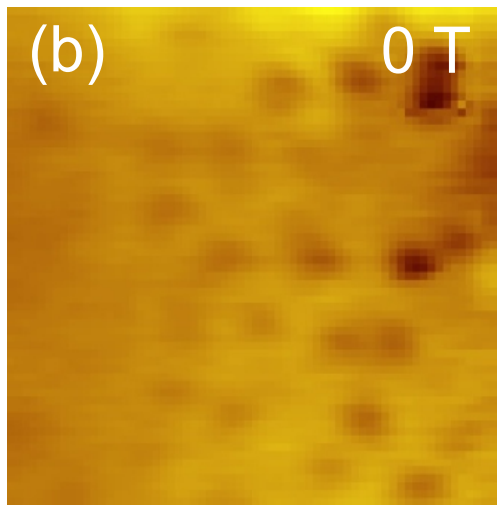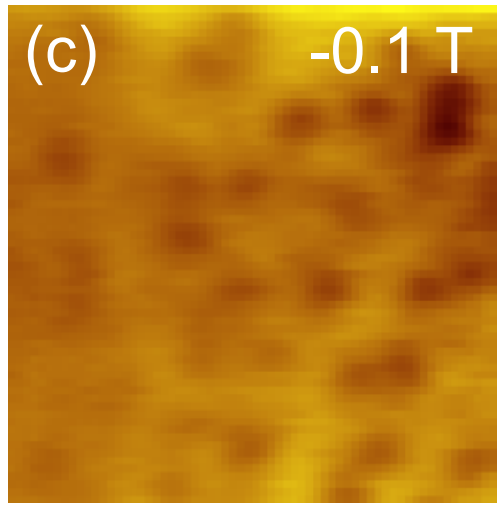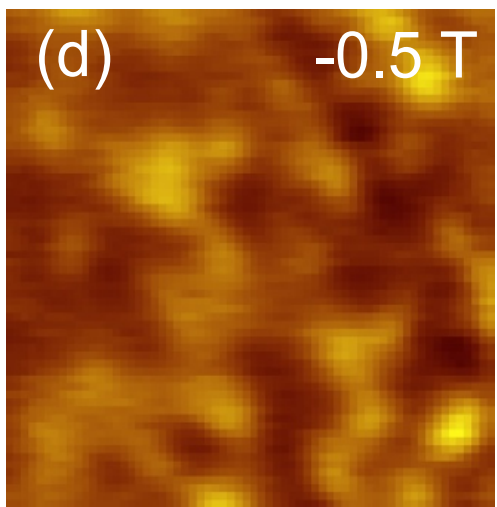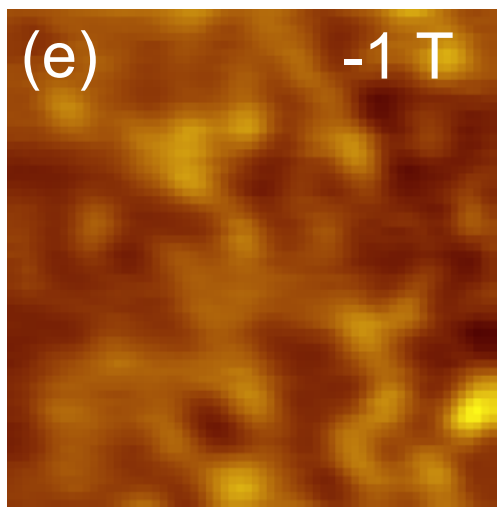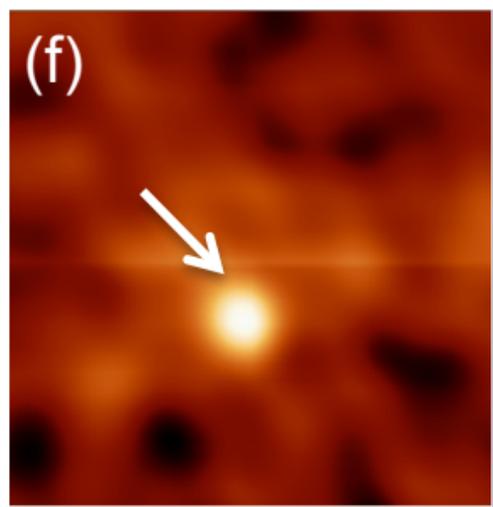

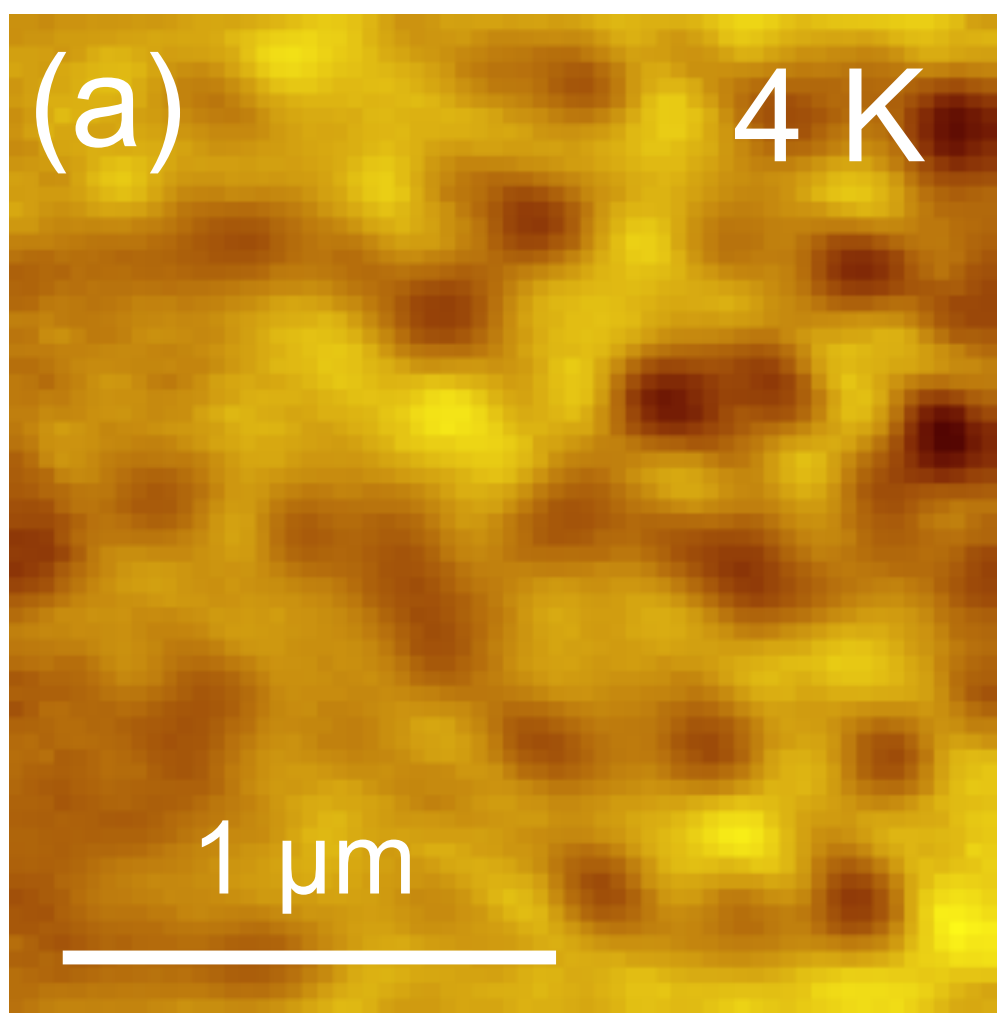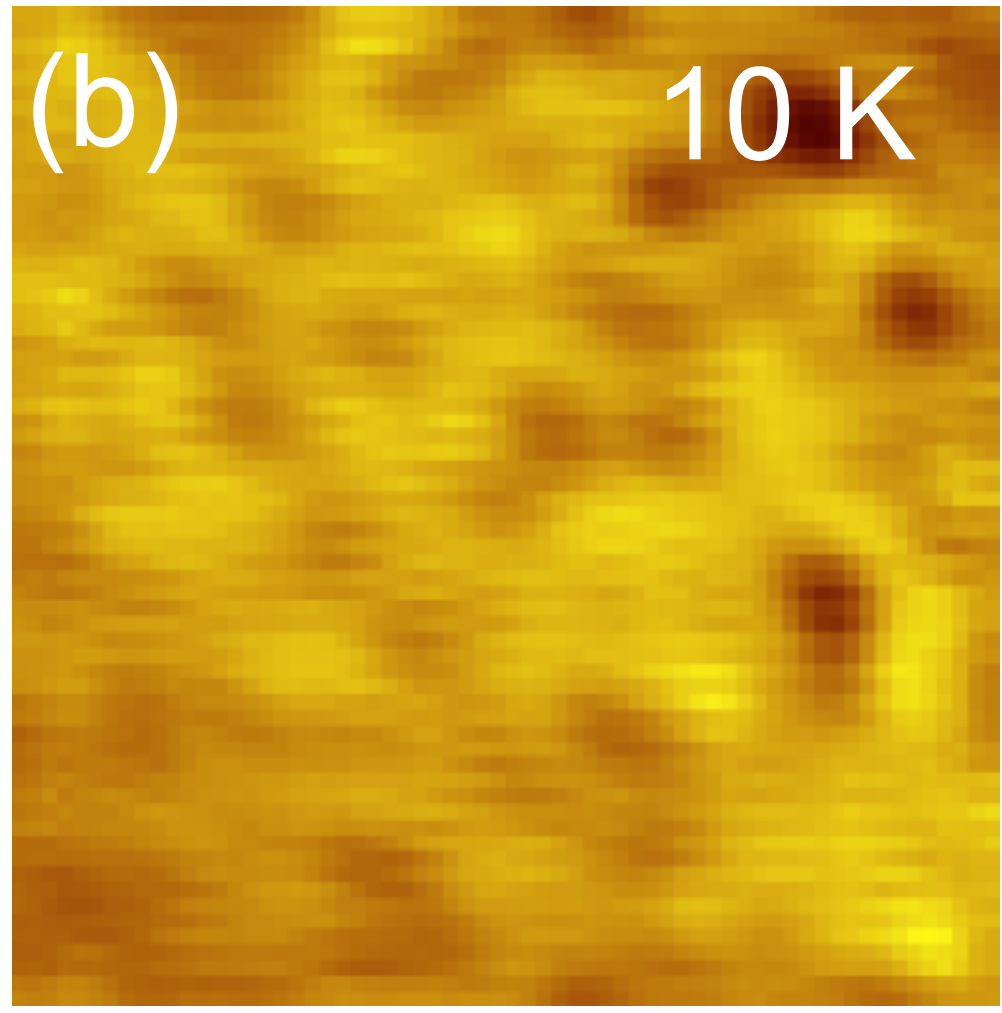